\def\be{\begin{equation}}
\def\ee{\end{equation}}
\def\ba{\begin{array}}
\def\ea{\end{array}}

\documentclass[prl,showpacs,twocolumn,amsmath]{revtex4}
\usepackage{pifont}
\usepackage{amssymb}
\def\qed{\leavevmode\unskip\penalty9999 \hbox{}\nobreak\hfill
     \quad\hbox{\leavevmode  \hbox to.77778em{%
               \hfil\vrule   \vbox to.675em%
               {\hrule width.6em\vfil\hrule}\vrule\hfil}}
     \par\vskip3pt}

\input amssym.def

\newtheorem{theorem}{Theorem}
\newtheorem{lemma}{Lemma}

\begin{document}

\title{\Large\bf Separability criteria based on Heisenberg-Weyl representation of density matrices}
\author{  Jingmei Chang$^1$, Meiyu Cui$^1$,  Tinggui Zhang$^{1,2\dag}$, Shao-Ming Fei$^{3,4}$\\[10pt]
\footnotesize
\small$1$ School of Mathematics and Statistics, Hainan Normal University,\\
\small Haikou 571158, P. R. China\\
\small$2$ Hainan Center for Mathematical Research, Hainan Normal
University,\\
\small Haikou 571158, P. R. China\\
$^3$Max-Planck-Institute for Mathematics in the Sciences,
 Leipzig 04103, Germany\\
$^4$School of Mathematical Sciences, Capital Normal University,
Beijing 100048, P. R. China\\
 \small $^\dag$
Correspondence to tinggui333@163.com}
\bigskip

\begin{abstract}
Separability is an important problem in theory of quantum entanglement.
By using the Bloch representation of quantum states in terms of the Heisenberg-Weyl observable
basis, we present a new separability criterion for bipartite quantum systems.
It is shown that this criterion can be better than the previous ones in detecting entanglement.
The results are generalized to multipartite quantum states.
\end{abstract}

 \maketitle

\bigskip

\section{Introduction}

Quantum entanglement is a fascinating phenomenon in quantum physics.
In recent decades, much works have been devoted to understand
entanglement as it plays important roles in many quantum information
processing. Nevertheless, there are still many problems remain
unsolved in the theory of quantum entanglement\cite{rpmk}. One basic
problem is to determine whether a given bipartite state is entangled
or separable. Although the problem is believed to be a
nondeterministic polynomial-time hard problem, there are a number of
operational criteria to deal with the problem, for example, the
positive partial transpose (PPT) criterion \cite{pera,mprh},
realignment criteria \cite{rudo,ckwl,Rudo,zzzg,aplc}, covariance
matrix criteria \cite{ghpo,gogo,gohp}, correlation matrix criteria
\cite{bblp,wmtm,SCB} and so on. More recently, some more
separability criteria have been proposed
\cite{LB,swlf,mlew,kfuj,fbwd,ISC}. Among them, Li et al. \cite{LB}
presented separability criteria based on correlation matrices and
the Bloch vectors of reduced density matrices. And by adding some
extra parameters, Ref.\cite{ISC} presents a more general
separability criterion for bipartite states in terms of the Bloch
representation of density matrices.

The state of two quantum systems A and B, acting on the finite-dimensional Hilbert space $H=H_{A}\otimes H_{B}$, is described by the density operator $\rho$.
A state $\rho$ is said to be separable if $\rho$ can be written as a convex combination of product vectors \cite{1rho}, i.e.
\begin{equation}\label{principal matrix}
\rho=\sum_{i}p_{i}|\psi_{i},\varphi_{i}\rangle\langle \psi_{i},\varphi_{i}|,
\end{equation}
where $0\leqslant p_{i}\leqslant 1$, $\sum_{i}p_{i}=1$,
and $|\psi_{i},\varphi_{i}\rangle=|\psi_{i}\rangle_{A}\otimes|\varphi_{i}\rangle_{B}$ ($|\psi\rangle_{A}\in H_{A}$ and $|\varphi\rangle_{B} \in H_{B}$).
The state $\rho$ is said to be entangled, when $\rho$ cannot be written as in form of Eq.(\ref{principal matrix}).

In this article, we put forward a new Bloch representation in terms of the Heisenberg-Weyl (HW) observable
basis \cite{2hw}. It is one of the standard Hermitian
generalization of Pauli operators, constructed from HW
operators \cite{rabp,smps,aacb,hxjn}. They have distinct properties
from those of Gell-Mann matrices \cite{2hw}, Based on the Heisenberg-Weyl representation of density matrices, we give a new separability criterion for
bipartite quantum states and multipartite states. By example, we show
that this criterion has advantages in determining whether a quantum state is separable or entangled.

\emph{HW observable basis}. First, we briefly
introduce the HW-operator basis  \cite{2hw}. The generalized Pauli ``phase" and ``shift" operators
are given by $Z=e^{\frac{i2\pi Q}{d}}$ and $X=e^{\frac{-i2\pi P}{d}}$, respectively,
$X|j\rangle=|j+1 $ mod $ d\rangle$ and $Z|j\rangle=e^{\frac{i2\pi
j}{d}}|j\rangle$. Q and P are the discrete position and momentum
operators describing a $d\times d$ grid.

The phase-space displacement operators for d-level systems are defined by
\begin{equation}\label{d definition}
\mathcal{D}(l,m)=Z^{l}X^{m}e^{\frac{-i\pi lm}{d}} \ ,
\end{equation}
i.e. \cite{rabp}
\begin{equation}\label{D definition}
\mathcal{D}(l,m)=\sum_{k=0}^{d-1}e^{\frac{2i\pi kl}{d}}|k\rangle\langle(k+m) mod\ d|,
\end{equation}
$\ l,m=0,1,\dots,d-1$. These non-Hermitian orthogonal basis operators satisfy the following orthogonality condition,
$Tr{\{\mathcal{D}(l,m)\mathcal{D}^{\dagger}(l',m')\}}=d\delta_{l,l^{'}}\delta_{m,m^{'}}$.

The complete set of Hermitian operators can be constructed from the HW operators $\mathcal{D}(l,m)$ by defining
\begin{equation}\label{Q definition}
\mathcal{Q}(l,m)=\mathcal{X}\mathcal{D}(l,m)+\mathcal{X^{*}}\mathcal{D}^{\dagger}(l,m) \ ,
\end{equation}
where ${\mathcal{X}}=(1\pm i)/2$.
$\mathcal{Q}(l,m)$ are the so called HW observable basis and satisfy the orthogonality condition,
\begin{equation}\label{Qtrace definition}
Tr{\{\mathcal{Q}(l,m)\mathcal{Q}(l',m')\}}=d\delta_{l,l^{'}}\delta_{m,m^{'}} \ .
\end{equation}

This basis simply reduces to the pauli matrices for $d=2$. When
$d=3$, we have
$$
 \mathcal{Q}(0,1)=\frac{1}{2}\left(
\begin{array}{ccc}
0&1+i&1-i\\
1-i&0&1+i\\
1+i&1-i&0
\end{array}
\right),$$$$ \mathcal{Q}(0,2)=\frac{1}{2}\left(
\begin{array}{ccc}
0&1-i&1+i\\
1+i&0&1-i\\
1-i&1+i&0
\end{array}
\right), $$$$ \mathcal{Q}(1,0)=\frac{1}{2}\left(
\begin{array}{ccc}
2&0&0\\
0&-1-\sqrt{3}&0\\
0&0&\sqrt{3}-1
\end{array}
\right),$$$$ \mathcal{Q}(1,1)=\frac{1}{\sqrt2}\left(
\begin{array}{ccc}
0&e^{\frac{\pi}{4}i}&e^{\frac{5}{12}\pi i}\\
e^{-\frac{\pi}{4}i}&0&e^{\frac{11}{12}\pi i}\\
e^{-\frac{5}{12}\pi i}&e^{-\frac{11}{12}\pi i}&0
\end{array}
\right), $$

$$\mathcal{Q}(1,2)=\frac{1}{\sqrt2}\left(
\begin{array}{ccc}
0&e^{-\frac{11}{12}\pi i}&e^{\frac{\pi}{4}i}\\
e^{\frac{11}{12}\pi i}&0&e^{\frac{5}{12}\pi i}\\
e^{-\frac{\pi}{4}i}&e^{-\frac{5}{12}\pi i}&0
\end{array}
\right),$$$$ \mathcal{Q}(2,0)=\frac{1}{2}\left(
\begin{array}{ccc}
2&0&0\\
0&\sqrt3-1&0\\
0&0&-1-\sqrt{3}
\end{array}
\right), $$$$ \mathcal{Q}(2,1)=\frac{1}{\sqrt2}\left(
\begin{array}{ccc}
0&e^{\frac{\pi}{4}i}&e^{-\frac{11}{12}\pi i}\\
e^{-\frac{\pi}{4}i}&0&e^{-\frac{5}{12}\pi i}\\
e^{\frac{11}{12}\pi i}&e^{\frac{5}{12}\pi i}&0
\end{array}
\right),$$$$
 \mathcal{Q}(2,2)=\frac{1}{\sqrt2}\left(
\begin{array}{ccc}
0&e^{\frac{5}{12}\pi i}&e^{\frac{\pi}{4}i}\\
e^{-\frac{5}{12}\pi i}&0&e^{-\frac{11}{12}\pi i}\\
e^{-\frac{\pi}{4}i}&e^{\frac{11}{12}\pi i}&0
\end{array}
\right).$$

\section{Bloch Representation under Heisenberg-Wely Observables}

A state $\rho\in \mathbb{C}^{d}$ of single quantum system can be
expressed in terms of the $d\times d$ identity operator $I_{d}$ and the $d^{2}-1$ traceless
Hermitian HW observable operators $\mathcal{Q}(l,m)$,
\begin{equation}\label{state definition}
\begin{array}{ll}
\rho&=\displaystyle{\frac{1}{d}}\sum_{l,m=0}^{d-1}r_{lm}\mathcal{Q}(l,m)\\[4mm]
&=\displaystyle\frac{1}{d}(I_{d}+\sum_{\substack{l,m=0
\\ (l,m)\ne (0,0)}}^{d-1}r_{lm}\mathcal{Q}(l,m)) \ ,
\end{array}
\end{equation}
where $\mathcal{Q}(0,0)=I_{d}$.
The coefficients $r_{lm}$ in Eq.(\ref{state definition}) are given by
\begin{equation}\label{trace definition}\nonumber
r_{lm}=Tr{(\rho\mathcal{Q}(l,m))} \ ,
\end{equation}
where $l=0,1,\cdots,d-1$, $m=0,1,\cdots,d-1$ and $(l,m)\ne (0,0)$. We denote
\begin{eqnarray}\label{rrr}\nonumber
\textbf{r}&=&(r_{0,1},r_{0,2},\cdots,r_{0,d-1},r_{1,0},\cdots,r_{1,d-1},\\
&&\cdots,r_{d-1,0},\cdots,r_{d-1,d-1}).
\end{eqnarray}

\begin{lemma}  For pure states,
\be\label{Fan definition}
\parallel\textbf{r}\parallel_{2}=\sqrt{d-1} \ ,
\ee
where $\parallel\cdot\parallel_{2}$ is the Euclidean norm on $\mathbb{R}^{d^2-1}$.
\end{lemma}

{\bf{Proof:}} According to the Eq.(\ref{state definition}), we have
\begin{eqnarray*}
Tr{\rho^2}&=& {\frac{1}{d^2}}\sum_{l,m,l',m'=0}^{d-1}r_{lm}r_{l'm'}Tr{\{\mathcal{Q}(l,m)\mathcal{Q}(l',m')\}} \\
           &=& {\frac{1}{d^2}} \cdot d \cdot \sum_{l,m=0}^{d-1}r_{lm}^2 \\
           &=& {\frac{1}{d}}(1+\sum_{\substack{l,m=0 \\ (l,m)\ne (0,0)}}^{d-1}r_{lm}^2)= {\frac{1}{d}}(1+\parallel\textbf{r}\parallel_{2}^2).
\end{eqnarray*}
Since $\rho$ is a pure state, one has $Tr{\rho^2}=Tr{\rho}=1$. Therefore
$\parallel\textbf{r}\parallel_{2}^2=d-1$.
$\hfill\Box$

Now consider bipartite states $\rho\in \mathbb{C}^{d_1}\otimes\mathbb{C}^{d_2}$. Any state $\rho$ can be similarly represented as \cite{2rho}
\begin{equation}\label{rho definition}
\begin{aligned}
\rho=& {\frac{1}{d_1d_2}}(I_{d_1}\otimes I_{d_2}+\sum_{\substack{(l,m)\ne
(0,0)}}r_{lm}\mathcal{Q}(l,m)\otimes I_{d_2} \\&+\sum_{(k,n)\ne
(0,0)}s_{kn}I_{d_1}\otimes\mathcal{\widetilde{Q}}(k,n) \\
&+\sum_{(l,m),(k,n)\ne
(0,0)}t_{lmkn}\mathcal{Q}(l,m)\otimes\mathcal{\widetilde{Q}}(k,n)),
\end{aligned}
\end{equation}
In particular,
\begin{equation}\label{AB definition}\nonumber
\begin{aligned}
r_{lm}&=Tr{\{\rho \mathcal{Q}(l,m))\otimes I_{d_2}\}}, \\
s_{kn}&=Tr{\{\rho I_{d_1}\otimes \mathcal{\widetilde{Q}}(k,n))\}}, \\
t_{lmkn}&=Tr{\{\rho \mathcal{Q}(l,m))\otimes \mathcal{\widetilde{Q}}(k,n)\}},
\end{aligned}
\end{equation}
where $l,m=0,\cdots,d_1-1$, $k,n=0,\cdots,d_2-1$ and $(l,m),(k,n)\ne (0,0)$.

\section{Separability criteria for bipartite states}

Similarly to (\ref{rrr}), we denote
$
\textbf{r}=(r_{0,1},\cdots,r_{0,d_1-1},\cdots,r_{d_1-1,0},
\cdots,r_{d_1-1,d_1-1})^{t}
$
and
$
\textbf{s}=(s_{0,1},\cdots,s_{0,d_2-1},\cdots,s_{d_2-1,0},
\cdots,s_{d_2-1,d_2-1})^{t}$, where $t$ stands for transpose.
Set $T=(t_{ij})$, where the entries $t_{ij}$ are given by the coefficients
$t_{lmkn}$, $l,m=0,\cdots,d_1-1$, $k,n=0,\cdots,d_2-1$, $(l,m),(k,n)\ne(0,0)$,
with the first two indices $lm$ associated with the array index $i$, and
the last two indices $kn$ with the column index $j$ of $T$.

Let us consider the following matrix,
\begin{equation}\label{sss}
\mathcal{S}_{\alpha,\beta}^{m}(\rho) = \left(
  \begin{array}{cc}
  \alpha\beta E_{m\times m} & \beta\omega_{m}(\textbf{s})^{t} \\
  \alpha\omega_{m}(\textbf{r}) & T
  \end{array}
\right),
\end{equation}
where
$$\omega_{m}(\textbf{x}) =
\underbrace{\left(
  \begin{array}{ccc}
  \textbf{x} & \cdots & \textbf{x}
  \end{array}
\right)}_{m \ columns},
$$
$\alpha$ and $\beta$ are nonnegative real numbers, $m$ is a given
natural number, $E_{m\times m}$ is an $m\times m$ matrix with all
entries being $1$. We have the following theorem.

\begin{theorem} If the state $\rho$ in $\mathbb{C}^{d_1}\otimes\mathbb{C}^{d_2}$ is separable, then
\be\label{theorem 1}
\parallel\mathcal{S}_{\alpha,\beta}^{m}(\rho)\parallel_{tr}\le\sqrt{(m\beta^2+d_1-1)(m\alpha^2+d_2-1)} \ ,
\ee
where $||\cdot||_{tr}$ stands for the trace norm (the sum of singular values).
\end{theorem}

{\bf{Proof:}} \ Since $\rho$ is separable, from \cite{SCB} there
exist vectors $\textbf{u}_{i}\in\mathbb{R}^{d_1^2-1}$,
$\textbf{v}_{i}\in\mathbb{R}^{d_2^2-1}$ satisfying Eq.(\ref{Fan
definition}), and weights $p_{i}$ satisfying $0\leq p_{i}\le 1$,
$\sum_{i}p_{i}=1$ such that
\begin{eqnarray*}\label{Tuv definition}
T=\sum_{i} p_{i}\textbf{u}_{i}\textbf{v}_{i}^{t} \ , \ \textbf{r}=\sum_{i} p_{i}\textbf{u}_{i} \ , \ \textbf{s}=\sum_{i} p_{i}\textbf{v}_{i}.
\end{eqnarray*}
From Lemma 1, we have
\begin{eqnarray*}
\parallel\textbf{u}_{i}\parallel_{2}=\sqrt{d_1-1} \ , \ \parallel\textbf{v}_{i}\parallel_{2}=\sqrt{d_2-1} \ .
\end{eqnarray*}
The matrix (\ref{sss}) has the form,
\begin{eqnarray*}
\mathcal{S}_{\alpha,\beta}^{m}(\rho) &=& \sum_{i}p_{i}\left(
  \begin{array}{cc}
  \alpha\beta E_{m\times m} & \beta\omega_{m}(\textbf{v}_{i})^{t} \\
  \alpha\omega_{m}(\textbf{u}_{i}) & \textbf{u}_{i}\textbf{v}_{i}^{t}
  \end{array}
\right), \\
&=& \sum_{i}p_{i}\left(
  \begin{array}{c}
  \beta E_{m\times 1} \\
  \textbf{u}_{i}
  \end{array}
\right)
\left(
  \begin{array}{cc}
  \alpha E_{1\times m} & \textbf{v}_{i}^{t}
  \end{array}
\right).
\end{eqnarray*}
Hence
\begin{eqnarray*}
\parallel\mathcal{S}_{\alpha,\beta}^{m}(\rho)\parallel_{tr} &=& \parallel\sum_{i}p_{i}\left(
  \begin{array}{c}
  \beta E_{m\times 1} \\
  \textbf{u}_{i}
  \end{array}
\right)
\left(
  \begin{array}{cc}
  \alpha E_{1\times m} & \textbf{v}_{i}^{t}
  \end{array}
\right)\parallel_{tr} \\
&\le& \sum_{i}p_{i}\parallel\left(
  \begin{array}{c}
  \beta E_{m\times 1} \\
  \textbf{u}_{i}
  \end{array}
\right)
\left(
  \begin{array}{cc}
  \alpha E_{1\times m} & \textbf{v}_{i}^{t}
  \end{array}
\right)\parallel_{tr}.
\end{eqnarray*}

Accounting to that for any vectors $|i\rangle=(i_1,i_2,\cdots,i_{m})^{t}$ and $|j\rangle=(j_1,j_2,\cdots,j_{n})^{t}$, one has
\begin{eqnarray}\label{ij}\nonumber
\parallel|i\rangle\langle j|\parallel_{tr} &=& Tr{\sqrt{(|i\rangle\langle j|)^{\dagger}|i\rangle\langle j|}} \\\nonumber
&=& Tr{\sqrt{|j\rangle(\langle i||i\rangle)\langle j|}} \\\nonumber
&=& \sqrt{(i_1^2+i_2^2+\cdots+i_m^2)(j_1^2+j_2^2+\cdots+j_n^2)} \\
&=& \parallel|i\rangle\parallel_{2}\parallel|j\rangle\parallel_{2}\ ,
\end{eqnarray}
we have
\begin{eqnarray*}
&&\sum_{i}p_{i}\parallel\left(
  \begin{array}{c}
  \beta E_{m\times 1} \\
  \textbf{u}_{i}
  \end{array}
\right)
\left(
  \begin{array}{cc}
  \alpha E_{1\times m} & \textbf{v}_{i}^{t}
  \end{array}
\right)\parallel_{tr} \\&=& \parallel\left(
  \begin{array}{c}
  \beta E_{m\times 1} \\
  \textbf{u}_{i}
  \end{array}
\right)
\parallel_{2}\parallel\left(
  \begin{array}{cc}
  \alpha E_{m\times 1} \\
  \textbf{v}_{i}
  \end{array}
\right)\parallel_{2} \\
&=& \sqrt{(m\beta^2+d_1-1)(m\alpha^2+d_2-1)} \ ,
\end{eqnarray*}
which gives rise to (\ref{theorem 1}).
$\hfill\Box$

{\sf Remark} Theorem 1 implies that
a pure bipartite quantum state in Bloch representation Eq. (\ref{rho definition}) is separable if and only if
\begin{eqnarray*}
\mathcal{S}_{\alpha,\beta}^{m}(\rho)&=&\left(
  \begin{array}{c}
  \beta E_{m\times 1} \\
  \textbf{r}
  \end{array}
\right) \left(
  \begin{array}{cc}
  \alpha E_{1\times m} & \textbf{s}^{t}
  \end{array}
\right)\\&=&\left(
  \begin{array}{cc}
  \alpha\beta E_{m\times m} & \beta\omega_{m}(\textbf{s})^{t} \\
  \alpha\omega_{m}(\textbf{r}) & T
  \end{array}
\right).
\end{eqnarray*}
Note that Eq.(\ref{rho definition}) can be rewritten as
\begin{eqnarray*}
\begin{aligned}
\rho=\rho_{A}\otimes\rho_{B}+{\frac{1}{d_1d_2}}[(t_{lmkn}-r_{lm}s_{kn})\mathcal{Q}(l,m)\otimes\mathcal{\widetilde{Q}}(k,n)],
\end{aligned}
\end{eqnarray*}
where $\rho_{A}$ and $\rho_{B}$ are the reduced density matrices.
Since $\mathcal{Q}(l,m)\otimes\mathcal{\widetilde{Q}}(k,n)$
are linearly independent,
$(t_{lmkn}-r_{lm}s_{kn})\mathcal{Q}(l,m)\otimes\mathcal{\widetilde{Q}}(k,n)=0$
if and only if $t_{lmkn}-r_{lm}s_{kn}=0$, i.e.
$T=\textbf{r}\textbf{s}^{t}$, for any $l,m,k,n$.
Moreover, for $d_1=d_2=2$, the HW observable basis is equivalent to Pauli matrices. In this case
Theorem 1 is equivalent to the Theorem 1 in \cite{ISC}.

For high dimensional quantum states, let us consider
the following $2\times4$ bound entangled state \cite{example} as an example,
$$
\rho=\frac{1}{7b+1}\left(
  \begin{array}{ccccccccc}
    b & 0 & 0 & 0 & 0 & b & 0 & 0 \\
    0 & b & 0 & 0 & 0 & 0 & b & 0 \\
    0 & 0 & b & 0 & 0 & 0 & 0 & b \\
    0 & 0 & 0 & b & 0 & 0 & 0 & 0 \\
    0 & 0 & 0 & 0 & \frac{1}{2}(1+b) & 0 & 0 & \frac{1}{2}\sqrt{1-b^2} \\
    b & 0 & 0 & 0 & 0 & b & 0 & 0 \\
    0 & b & 0 & 0 & 0 & 0 & b & 0 \\
    0 & 0 & b & 0 & \frac{1}{2}\sqrt{1-b^2} & 0 & 0 & \frac{1}{2}(1+b) \\
  \end{array}
\right).
$$
where $0<b<1$. We mix the above state with state $|\xi\rangle=\frac{1}{\sqrt{2}}(|00\rangle+|11\rangle)$,
$$
\rho_{x}=x|\xi\rangle\langle\xi|+(1-x)\rho.
$$
By choosing
$$
\alpha=\frac{1}{2},\ \ \beta=\sqrt{\frac{2}{11}},\ \ m=1,\ \ b=0.9,
$$
our Theorem 1 can detect the entanglement in $\rho_{x}$ for $  \
0.2234\le x \le 1$, while the Theorem 1 in \cite{ISC}, the V-B criterion
\cite{SCB} and the L-B criterion \cite{LB} can only detect the
entanglement in $\rho_{x}$ for $0.2320\le x\le 1$, $0.2293\le x\le
1$ and $0.2841\le x\le 1$, respectively. In this case, our criterion
is better in detecting entanglement.

Here, instead of (\ref{Q definition}), if we define
$\mathcal{Q}(l,m)=k(\mathcal{X}\mathcal{D}(l,m)+\mathcal{X^{*}}\mathcal{D}^{\dagger}(l,m))$,
$k=\sqrt{{2}/{d}}$, then $\parallel\textbf{r}\parallel_{2}=\sqrt{\frac{d(d-1)}{2}}$,
and the conclusion becomes
$$
\parallel\mathcal{S}_{\alpha,\beta}^{m}(\rho)\parallel_{tr}\le\frac{1}{2}\sqrt{(2m\beta^2+d_1^2-d_1)(2m\alpha^2+d_2^2-d_2)} \ .
$$
In this case, the least upper bound in Theorem 1 is equal to the Theorem 1 \cite{ISC}.

\section{Separability criteria for multipartite states}

We now generalize our result in Theorem 1 to multipartite case.
Let $\mathcal{S}$ be an $f_1\times\dots\times f_N$ tensor, $A$ and
$\bar{A}$ be two nonempty subsets of $\{1,\dots,N\}$ satisfying
$A\cup\bar{A}={1,\dots,N}$. Let $\mathcal{S}^{A|\bar{A}}$ denote
the $A$, $\bar{A}$ matricization of $\mathcal{S}$, see
\cite{matricization,ISC} for detail.

For any state $\rho$ in
$\mathbb{C}^{d_1}\otimes\dots\otimes\mathbb{C}^{d_N}$, we import a
natural number $m$ and nonnegative real parameters
$\alpha_1,\dots\alpha_N$, and define
\begin{equation}\label{delta definition}\nonumber
\delta_{(k_i,n_i)}^{(d_i)}=\left\{
\begin{aligned}
 \alpha_{i}I_{d_i}, \ \ \ \ \ \ \ \ \ \ \ \ \ \ \  \ \ \ \ \ & 1\le k_i=n_i\le m \\[4mm]
 Q^{(d_i)}(k_i-m,n_i-m),\ &\begin{array}{l} m\le k_i\le d_i+m-1,\\
 m\le n_i\le d_i+m-1,\\
 (k_i,n_i)\ne (m,m),\end{array}
\end{aligned}
\right.
\end{equation}
where $i=1,\dots,N$ and $Q^{(d_i)}(k,n)$ be the traceless Hermitian
HW observable basis and satisfy the orthogonality relation
$Tr{\{\mathcal{Q}^{(d_i)}(k,n)\mathcal{Q}^{(d_i)}(k',n')\}}=d_i\delta_{l,l^{'}}\delta_{m,m^{'}}$.
Denote $\mathcal{W}^{(m)}_{\alpha_1,\alpha_2,\dots,\alpha_N}(\rho)$
the tensor given by elements of the following form,
\begin{equation}\label{w definition}\nonumber
w_{(k_1,n_1)\dots
(k_N,n_N)}=Tr{(\rho\delta^{(d_1)}_{(k_1,n_1)}\otimes\dots\otimes\delta^{(d_N)}_{(k_N,n_N)})},
\end{equation}
where $1\le k_i=n_i\le m$ and $m\le k_i,n_i\le d_i+m-1$.
Below we give the full separability
criterion based on
$\mathcal{W}^{(m)}_{\alpha_1,\alpha_2,\dots,\alpha_N}(\rho)$.

\begin{theorem} If a state $\rho$ in $\mathbb{C}^{d_1}\otimes\dots\otimes\mathbb{C}^{d_N}$ is fully separable, then for any subset $A$ of $\{1,\dots,N\}$, we have \be\label{multilinear Fan relation}
\parallel(\mathcal{W}^{(m)}_{\alpha_1,\alpha_2,\dots,\alpha_N}(\rho))^{A|\bar A}\parallel_{tr}\le\prod_{k=1}^{N}\sqrt{(m\alpha_k^2+d_k-1)}.
\ee
\end{theorem}

{\bf{Proof:}} Without loss of generality, we assume
\begin{eqnarray*}
A&=&\{q_1,\dots,q_M\},~~~~~~~\ q_1<\dots<q_M, \\
\bar{A}&=&\{q_{M+1},\dots,q_N\},~~~\ q_M+1<\dots<q_N.
\end{eqnarray*}
Since $\rho$ is fully separable, from \cite{fullysep} there
exist vectors $\textbf{u}^{(k)}_i\in\mathbb{R}^{d^2_k-1}$ such that
\begin{eqnarray*}
\mathcal{W}^{(m)}_{\alpha_1,\alpha_2,\dots,\alpha_N}(\rho)=\sum_i
p_i\left(
  \begin{array}{c}
  \alpha_1 E_{m\times 1} \\
  \textbf{u}_{i}^{(1)}
  \end{array}
\right)\otimes\dots\otimes\left(
  \begin{array}{c}
  \alpha_N E_{m\times 1} \\
  \textbf{u}_{i}^{(N)}
  \end{array}
\right),
\end{eqnarray*}
where $\parallel\textbf{u}_{i}^{(k)}\parallel_2=\sqrt{d_k-1}$.
Thus
\begin{equation*}
\begin{aligned}
& \ \ \
\parallel(\mathcal{W}^{(m)}_{\alpha_1,\alpha_2,\dots,\alpha_N}(\rho))^{A|\bar A}\parallel_{tr} \\
&=\parallel\sum_i p_i\mathop\otimes\limits_{l=1}^{M}\left(
  \begin{array}{c}
  \alpha_{ql} E_{m\times 1} \\
  \textbf{u}_{i}^{(q_l)}
  \end{array}
\right)\mathop\otimes\limits_{p=M+1}^{N}\left(
  \begin{array}{c}
  \alpha_{q_p} E_{m\times 1} \\
  \textbf{u}_{i}^{(q_p)}
  \end{array}
\right)^t\parallel_{tr}\\
&\le\sum_ip_i\parallel\mathop\otimes\limits_{l=1}^{M}\left(
  \begin{array}{c}
  \alpha_{q_l} E_{m\times 1} \\
  \textbf{u}_{i}^{(q_l)}
  \end{array}
\right)\mathop\otimes\limits_{p=M+1}^{N}\left(
  \begin{array}{c}
  \alpha_{q_p} E_{m\times 1} \\
  \textbf{u}_{i}^{(q_p)}
  \end{array}
\right)^t\parallel_{tr}\\
&=\sum_ip_i\parallel\mathop\otimes\limits_{l=1}^{M}\left(
  \begin{array}{c}
  \alpha_{q_l} E_{m\times 1} \\
  \textbf{u}_{i}^{(q_l)}
  \end{array}
\right)\parallel_2
\parallel\mathop\otimes\limits_{p=M+1}^{N}\left(
  \begin{array}{c}
  \alpha_{q_p} E_{m\times 1} \\
  \textbf{u}_{i}^{(q_p)}
  \end{array}
\right)\parallel_{2} \\
&=\sum_ip_i\sqrt{tr(\mathop\otimes\limits_{l=1}^{M}(\left(
  \begin{array}{c}
  \alpha_{q_l} E_{m\times 1} \\
  \textbf{u}_{i}^{(q_l)}
  \end{array}
\right)^{\dagger}\left(
  \begin{array}{c}
  \alpha_{q_l} E_{m\times 1} \\
  \textbf{u}_{i}^{(q_l)}
  \end{array}
\right)))} \\
& \ \ \ \ \ \cdot\sqrt{tr(\mathop\otimes\limits_{p=M+1}^{N}(\left(
  \begin{array}{c}
  \alpha_{q_p} E_{m\times 1} \\
  \textbf{u}_{i}^{(q_p)}
  \end{array}
\right)^{\dagger}\left(
  \begin{array}{c}
  \alpha_{q_p} E_{m\times 1} \\
  \textbf{u}_{i}^{(q_p)}
  \end{array}
\right)))} \\
&=\sum_ip_i\sqrt{\sum_{k=1}^{N} tr(\left(
  \begin{array}{c}
  \alpha_{k} E_{m\times 1} \\
  \textbf{u}_{i}^{(k)}
  \end{array}
\right)^{\dagger}\left(
  \begin{array}{c}
  \alpha_{k} E_{m\times 1} \\
  \textbf{u}_{i}^{(k)}
  \end{array}
\right))} \\
&=\prod_{k=1}^{N}\sqrt{(m\alpha_k^2+d_k-1)},
\end{aligned}
\end{equation*}
where we have used the equality Eq.(\ref{ij}) and $tr{(A\otimes B)}=tr{A}\cdot tr{B}$.

\section{conclusion}

We have studied the separability problem based on the Bloch representation of density matrices
in terms of the Heisenberg-Weyl observable basis.
New separability criteria have been derived for both bipartite and multipartite quantum systems,
which provide more efficient ways in detecting quantum entanglement for certain kinds of quantum states.
These criteria can experimentally implemented.
The results may highlight approaches in dealing with the separability problem by using
suitable observable basis.

\bigskip
\noindent{\bf Acknowledgments}\ We thank Shu-Qian Shen for useful
discussions. The work is partly supported by the NSF of China under
Grant No. 11501153, No. 11661031 and No. 11675113; the NSF of Hainan Province under
Grant No. 20161006.

\end{document}